\documentclass[%
aip,
 amsmath,amssymb,
 reprint,%
]{revtex4-1}

\usepackage{graphicx}
\usepackage{dcolumn}
\usepackage{bm}
\usepackage{bbold}
\usepackage{amsmath,amsfonts,amsthm,amssymb}
\usepackage[bookmarksnumbered,pdfpagelabels=true,plainpages=false,colorlinks=true,linkcolor=blue,citecolor=blue,urlcolor=blue]{hyperref}
\usepackage[normalem]{ulem}

\newcommand\footnoteref[1]{\protected@xdef\@thefnmark{\ref{#1}}\@footnotemark}

\newcommand{\pt}{\partial}
\newcommand{\mb}{\mathbf}
\newcommand{\mc}{\mathcal}

\usepackage[utf8]{inputenc}
\usepackage[T1]{fontenc}
\usepackage{mathptmx}
\usepackage{etoolbox}

\makeatletter
\def\@email#1#2{%
 \endgroup
 \patchcmd{\titleblock@produce}
  {\frontmatter@RRAPformat}
  {\frontmatter@RRAPformat{\produce@RRAP{*#1\href{mailto:#2}{#2}}}\frontmatter@RRAPformat}
  {}{}
}%
\makeatother
\begin{document}

\preprint{AIP/123-QED}

\title{Skyrmion Qubits:\\Challenges For Future Quantum Computing Applications}

\author{Christina Psaroudaki}
\affiliation{Laboratoire de Physique de l’\'{E}cole normale sup\'{e}rieure, ENS, Universit\'{e} PSL, CNRS, Sorbonne Universit\'{e}, Universit\'{e} de Paris, F-75005 Paris, France}
\email{christina.psaroudaki@phys.ens.fr}

\author{Elias Peraticos}
\affiliation{Division of Physics and Applied Physics, School of Physical and Mathematical Sciences, Nanyang Technological University, 21 Nanyang Link 637371, Singapore}

\author{Christos Panagopoulos}
\affiliation{Division of Physics and Applied Physics, School of Physical and Mathematical Sciences, Nanyang Technological University, 21 Nanyang Link 637371, Singapore} 
\email{christos@ntu.edu.sg}

\date{\today}

\begin{abstract}
Magnetic nano-skyrmions develop quantized helicity excitations, and the quantum tunneling between nano-skyrmions possessing distinct helicities is indicative of the quantum nature of these particles. Experimental methods capable of non-destructively resolving the quantum aspects of topological spin textures, their local dynamical response, and their functionality now promise practical device architectures for quantum operations. With abilities to measure, engineer, and control matter at the atomic level, nano-skyrmions present opportunities to translate ideas into solid-state technologies. Proof-of-concept devices will offer electrical control over the helicity, opening a promising new pathway towards functionalizing collective spin states for the realization of a quantum computer based on skyrmions. This Perspective aims to discuss developments and challenges in this new research avenue in quantum magnetism and quantum information.
\end{abstract}

\maketitle
\noindent Authors to whom correspondence should be addressed: Christina Psaroudaki, christina.psaroudaki@phys.ens.fr; Christos Panagopoulos, christos@ntu.edu.sg

\section{Introduction}

Quantum computers have the potential to revolutionize data storage and processing beyond their conventional counterparts \cite{nielsen_chuang_2010}, making it a prominent area of research. Although quantum computers hold great promise, their realization and the identification of practical applications pose several challenges \cite{PRXQuantum.2.017001}. Research and development of new qubit technologies are therefore pursued intensely across different solid-state platforms. Among them, topological spin textures, such as magnetic skyrmions, are emerging as promising macroscopic qubits \cite{PhysRevLett.127.067201} due to their topological stability and nanosize. These topological excitations with particle-like behavior are highly resilient to external perturbations \cite{Fert2017,doi:10.1021/acs.chemrev.0c00297} and have primarily been exploited for a wide range of classical applications, ranging from spintronics devices \cite{Fert2013,Fert2017} to unconventional computation platforms \cite{PhysRevApplied.9.064018,Song2020}. 

The observation of nanometer-scale skyrmions \cite{Heinze2011,doi:10.1126/science.aau0968,Hirschberger2019} along with their stability in the milli-Kelvin regime, inspired an increasing number of studies on their quantum properties \cite{PhysRevX.7.041045,PhysRevB.94.134415,Wieser_2017,PhysRevB.88.060404,PhysRevResearch.2.033491,PhysRevB.92.245436,PhysRevB.103.224423,PhysRevB.98.024423,PhysRevLett.124.097202,Vlasov_2020, doi:10.1142/9789811231711_0004,Vlasov_2020,PhysRevB.103.L060404, PhysRevB.105.224416,PhysRevResearch.4.023111, PhysRevX.9.041063,PhysRevResearch.4.L032025}, expanding their potential for information processing from the classical to the quantum regime. These developments create the possibility of utilizing the quantum dynamics of magnetic skyrmions and bridge the skyrmionics field to quantum information, analogously to the highly interdisciplinary field of quantum magnonics \cite{YUAN20221}.  At the same time, the advancement of sensors capable of detecting magnetic signals with quantum sensitivity \cite{Rugar2004,Quirion2020} and the discovery of novel skyrmion hosting materials \cite{doi:10.1126/science.aau0968,Hirschberger2019} make the investigation of the quantum aspects of magnetic skyrmions experimentally feasible. 

\begin{figure*}[t]
	\centering
	\includegraphics[scale=0.4]{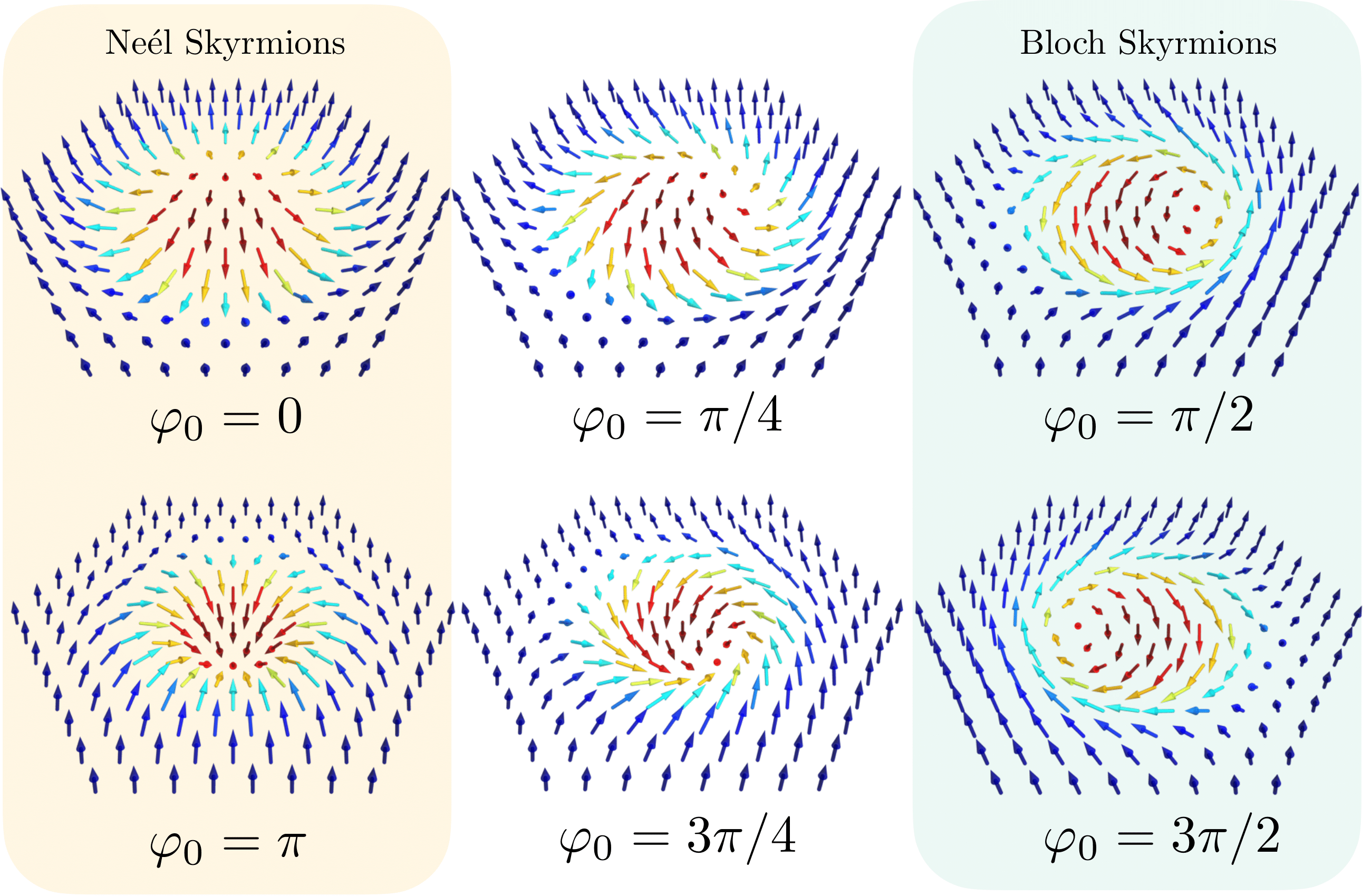}
	\caption{Illustrations of a series of 2D skyrmion textures with topological charge $Q=1$ and different helicities $\varphi_0$. The arrows denote the spin direction and the out-of-plane spin component is represented by color. Ne\'{e}l (left panel) and Bloch (right panel) are stabilized in DMI systems, while in frustrated magnets all possible configurations are allowed, separated by small energy gaps generated by additional anisotropies or dipolar interactions.}
	\label{Fig:multi_skyrmion}
\end{figure*}

Harnessing quantum functionalities requires identifying systems with discrete energy levels while satisfying several criteria, including scalability, controllability, and readout by nonvolatile techniques \cite{DiVincenzo2000}.  Skyrmion qubits use the energy-level quantization of skyrmion helicity to encode quantum information \cite{PhysRevLett.127.067201}. Thus, in contrast to other platforms based on natural microscopic systems, skyrmion qubits are built upon macroscopic collective variables, similar in nature to superconducting quantum circuits \cite{Clarke2008,doi:10.1126/science.1231930}.  The applicability of the classical variable of helicity to quantum operations depends on the feasibility of quantum tunneling between two macroscopic states with distinct helicities. For a nanoskyrmion that is sufficiently decoupled from its environment, tunable macroscopic quantum tunneling has been predicted to occur within experimental reach \cite{PhysRevB.106.104422}. 

Centrosymmetric materials are a prominent platform for constructing skyrmion quantum processors \cite{PhysRevLett.127.067201}.  In this class of materials, geometrical frustration renders the skyrmion helicity a quantum degree of freedom and leads to a higher density of skyrmions\cite{PhysRevB.93.064430}.  Here, skyrmions are considerably smaller \cite{doi:10.1021/acs.chemrev.0c00297} than those in non-centrosymmetric magnets.  Skyrmion lattice phases have already been observed in gadolinium compounds and perovskite oxides\cite{doi:10.1021/acs.chemrev.0c00297}.  Whereas new materials are expected to emerge as the field of frustrated magnetism progresses \cite{Batista2016}. 

Skyrmion qubits inherit the appealing physical properties of classical magnetic skyrmions, allowing for compact, high-density, and low-energy devices.  Their size, positional stability, lifetime, and energetics can be tuned by material engineering and geometrical control \cite{Vakili2021}. Their precise nucleation, detection, and dynamics have been extensively explored using a wide range of experimental methods \cite{PanagopoulosFinocchio}. Furthermore, the low energy-level spectra of skyrmion qubits are configurable by external magnetic and electric fields and can be designed to exhibit the desired qubit properties, including operation regimes, transition frequencies, anharmonicity, and qubit lifetime.  Microwaves can be utilized for qubit manipulation and gate operation, while nonvolatile measurement schemes can be employed for a reliable qubit state readout. Scalability is addressed by tailored coupling schemes between individual qubits.  

Despite the potential for competitive technology, the development of a skyrmion quantum processor will face near-term challenges.  On a fundamental level, these include among others the identification of candidate frustrated magnets with low damping, the microscopic understanding and control of noise sources, the design of optimal gate implementations, longer qubit coherence times, and tunable coupling of spatially separated qubits. On a device-specific level, the deterministic and precise skyrmion nucleation, the efficient determination of the fidelity of quantum operations, and robust and reproducible device fabrication are among the pressing issues to be addressed. 

The purpose of this Perspective is to present the quantum aspects of magnetic skyrmions and their usage in quantum operations and to discuss challenges and future directions in achieving skyrmion-based quantum technology.  The main advantage lies in the high degree of control in skyrmion manipulation, qubit parameter tunability, and all-magnetic device integration.  Whilst technical challenges would need to be overcome first, the application of skyrmions in the quantum regime is expected to lead to disruptive technologies, opening up new research opportunities in the fields of skyrmionics and quantum magnetism.

\section{Quantum Properties of Magnetic Skyrmions}

Magnetic skyrmions emerge as topologically nontrivial configurations of the magnetization field $\mb{m}(\mb{r},t)$ in certain helimagnetic materials \cite{doi:10.1021/acs.chemrev.0c00297}, characterized by a finite topological charge $Q=1/4\pi\int d\mb{r} \mathbf{m} \cdot(\pt_x \mb{m} \times \pt_y \mb{m})$ \cite{PAPANICOLAOU1991425}. They have been discovered in a plethora of magnetic bulk crystals \cite{doi:10.1021/acs.chemrev.0c00297} and multilayers \cite{JIANG20171} and extensively investigated theoretically and experimentally \cite{PanagopoulosFinocchio}.

Figure \ref{Fig:multi_skyrmion} shows a schematic of a series of single skyrmion configurations in two dimensions (2D). They are characterized by $\Phi = \mu \phi + \varphi_0$ and $\Theta= \Theta(\rho)$ with boundary conditions $\Theta(0)= \pi$ and $\Theta(\infty) = 0$, where $\{\rho,\phi\}$ are the polar coordinates, and $\mb{m} = [\sin \Theta \cos \Phi, \sin \Theta \sin \Phi, \cos \Theta]$ is the normalized magnetization. Here $\varphi_0$ denotes the skyrmion helicity and corresponds to the angle of the global rotation around the z-axis, while $\mu$ corresponds to the skyrmion vorticity and defines the topological charge, $Q=\mu$, for fixed boundary conditions. Due to the macroscopic size of magnetic skyrmions, their dynamics are typically governed by the purely classical Landau-Lifshitz-Gilbert (LLG) equation \cite{Landau:437299}.  The long-time dynamics of skyrmions reduce to a much simpler problem by considering generalized collective coordinates with long relaxation times \cite{PhysRevLett.100.127204}. They describe the overall movement of the magnetic texture and are associated with a continuous symmetry broken by the skyrmion configuration. 

For skyrmions at the nanometer scale and at temperatures much lower than the ordering transition, these degrees of freedom are expected to be affected by quantum fluctuations and are promoted to quantum mechanical operators \cite{doi:10.1142/S0217979219300056}. Their dynamics resemble quantum mechanical particles, which exhibit quantum tunneling between classically stable magnetic configurations, a necessary property to device elements of quantum computers \cite{DiVincenzo95}. The skyrmion center operator $\mb{\hat{R}}$ satisfies the commutation relation $[\hat{R}_x,\hat{R}_y] = i a^2/4\pi S Q$, with $a$ the lattice constant and $S$ the spin magnitude.  Here the magnetic length $\ell = a/\sqrt{4 \pi S \vert Q \vert }$ measures the extension of quantum fluctuations in the phase space of collective coordinates \cite{doi:10.1142/S0217979219300056}.  In the classical limit of large $S \rightarrow \infty$, it holds $\ell \gg a$, and quantum effects become weak.  The quantum treatment of the skyrmion propagation in chiral magnetic insulators revealed that observable quantum behavior may arise for $\mb{\hat{R}}$ in the presence of pinning defects that split the lowest Landau level into quantized levels \cite{PhysRevB.88.060404}.  In addition, the quantum fluctuations around the skyrmionic configuration give rise to a quantum mass term $\mc{M}$ with an explicit temperature dependence which remains finite even at a vanishingly small temperature \cite{PhysRevX.7.041045}.  For sufficiently small skyrmions, a quantum liquid phase appears as an intermediate phase between the skyrmion crystal and the ferromagnet \cite{PhysRevB.94.134415}.  

\begin{figure}
	\centering
	\includegraphics[width=1\linewidth]{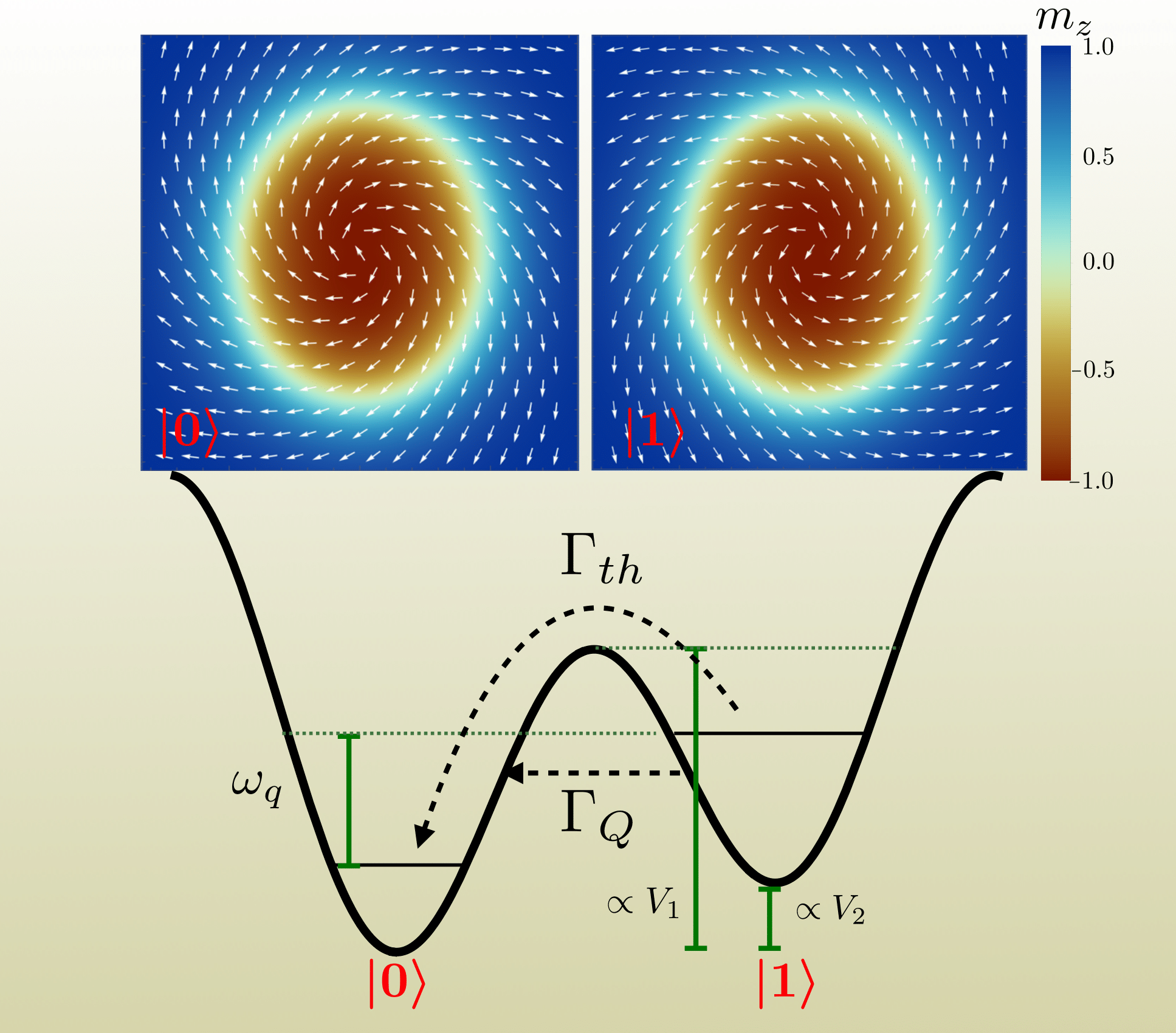}
	\caption{Representation of skyrmion helicity potential landscape. The logic space $\vert 0 \rangle$ and $\vert 1 \rangle$ is consisted of two macroscopic states with distinct helicities, localized at the two potential minima, separated by a tunable energy barrier. In the absence of bias $V_2$, macroscopic quantum tunneling through the barrier hybridizes the degenerate levels and introduces a level splitting. }
	\label{Fig:skyrmion_qubit}
\end{figure}

Evidence of quantum behavior in magnetic systems stems from the prediction and observation of macroscopic quantum tunneling events \cite{PhysRevLett.66.2802,PhysRevB.56.8129,doi:10.1126/science.284.5411.133,Thomas1996,Brooke2001}.  The quantum tunneling probability $\Gamma_Q$ is governed by the zero-temperature WKB exponent,  $\Gamma_Q \propto e^{-S_0}$,  with $S_0$ temperature-independent.  Above a critical temperature $T_c \approx \hbar U_0/ k_B S_0$, with $U_0$ the tunneling barrier, classical thermal events with $\Gamma_{th} \propto e^{-U_0/T}$ dominate over the quantum tunneling-induced transitions \cite{chudnovsky_tejada_1998} (see Fig.~\ref{Fig:skyrmion_qubit} for a schematic illustration). The quantum depinning of the skyrmion position out of an impurity potential suggests that sufficiently small magnetic skyrmions will behave as quantum objects with $T_c$ in the millikelvin temperature regime and $\Gamma_Q^{-1}$ within a few seconds \cite{PhysRevLett.124.097202}.  The problem resembles quantum depinning in the Hall-type dynamics of a vortex in high-$T_c$ superconductors \cite{RevModPhys.66.1125} or a charged spin texture in quantum Hall systems \cite{PhysRevLett.83.1411}, and describes the quantum behavior observed in similar mesoscopic particles \cite{Brooke2001,PhysRevB.85.180401}.  The quantum collapse of a classically stable skyrmion has been predicted to occur through under-barrier quantum tunneling to the ferromagnetic state \cite{PhysRevB.98.024423}, with $T_c$ estimated to be a few Kelvin for realistic material parameters \cite{Vlasov_2020}.  In the opposite scenario, single skyrmions are quantum mechanically nucleated from the ferromagnetic state in the presence of local magnetic fields \cite{doi:10.1142/9789811231711_0004}. 

In addition to the usual translational modes $\mb{R}$, a skyrmion stabilized in a model with an unbroken rotational symmetry possesses the skyrmion helicity $\varphi_0$ as an extra degree of freedom. 
Following the method of collective coordinate quantization, $\varphi_0$ and its conjugate momentum associated with global spin rotations $S_z = \int d\mb{r} (1-\cos\Theta) \pt_\phi \Phi$ are promoted to quantum operators satisfying $[\hat{\varphi}_0,\hat{S}_z] =i/S$ with the classical limit recovered when $S \rightarrow \infty$ \cite{PhysRevLett.127.067201}. Potential engineering achieved by introducing external perturbations for the skyrmion helicity allows the observation of diverse macroscopic quantum phenomena. For sufficiently small skyrmions with radius $\lambda \approx 5-10$ nm and a modest choice of effective spin $S \approx 1-10$, quantum tunneling processes between two macroscopic states with distinct helicities occur with $\Gamma_Q^{-1}$ within seconds below 100 mK. 
Macroscopic quantum coherence between degenerate states causes MHz-level energy tunnel splitting while tunneling in periodic potentials results in destructive quantum interference among equivalent tunneling paths \cite{PhysRevB.106.104422}. Hence, quantum tunneling, a fundamental aspect of quantum computing, is expected to be common and experimentally feasible in magnetic nano-skyrmions. 

\section{Skyrmion Quantum Computing}

Magnetic skyrmions are particularly promising candidates for information processing and computing due to their single-particle properties. In conventional logic devices, logic numbers 0 and 1 are associated with the presence or absence of magnetic skyrmions in racetrack devices \cite{10.1063/5.0042917,PhysRevApplied.15.064004}. Their physical properties, including nano-size, parameter tunability, and low-energy operation, render them potential candidates for unconventional computing approaches such as neuromorphic \cite{Grollier2020,Song2020}, reservoir \cite{Raab2022} and probabilistic \cite{PhysRevApplied.9.064018}. As discussed earlier, although magnetic skyrmions are macroscopic in size, under appropriate conditions they display generic quantum properties such as quantized energy levels, superposition of states, entanglement, and quantum tunneling. Thus, they offer a new class of primitive building blocks for quantum computers, extending their suitability from the classical to the quantum regime. 

\subsection{Skyrmion Qubits}

Skyrmions can be generated in magnetic systems through various mechanisms, often acting simultaneously. In systems that lack inversion symmetry, the relativistic DMI energetically stabilizes skyrmions \cite{BOGDANOV1994255}, with uniquely defined $\mu=\pm 1$ and a potential term $\sin(\varphi_0)$ (Bloch skyrmions) or $\cos(\varphi_0)$ (N\'eel skyrmions) depending on the DMI form, which in turn is determined by the crystal symmetry of the material (see Fig.~\ref{Fig:multi_skyrmion} for a schematic illustration). The DMI strength controls the skyrmion size, which typically ranges between 10-100 nm. Long-ranged magnetic dipolar interactions generate skyrmions of the order of 100 nm-1 $\mu$m with two degenerate lowest-energy states $\varphi_0=\pm \pi/2$ \cite{Nagaosa2013}. Frustrated exchange interactions \cite{Leonov2015} and four-spin exchange interactions \cite{Heinze2011} support atomic-sized skyrmion structures, and helicity can take any arbitrary value. Hence, magnetic frustration offers advantages for quantum information processing based on skyrmions because i) it provides a path to scale down the skyrmion size, a prerequisite for quantum effects to be strong and ii) skyrmion helicity is a zero mode and does not experience large potential terms. Importantly, helicity couples effectively to external perturbations, and the system can be driven to the optimum quantum regime \cite{PhysRevB.106.104422}. 

A skyrmion qubit is constructed based on the energy-level quantization of the helicity degree of freedom of a skyrmion stabilized in a 2D nanodisk. The reduced dynamics of the system are given in terms of the quantum Hamiltonian $H=\hat{S}_z^2/2\mc{M}+h_1 \hat{S}_z + V(\hat{\varphi}_0)$, where $\mc{M}$ is the mass for the skyrmion helicity, $h_1$ is the Zeeman term under the application of a uniform magnetic field, while the helicity potential can be written in its most general form as $V(\hat{\varphi}_0)= V_0 \cos 2 \hat{\varphi}_0 -V_1 \cos \hat{\varphi}_0 +V_2 \sin \hat{\varphi}_0$. The first term can be the result of dipole-dipole interaction \cite{Zhang2017}, in-plane uniaxial \cite{PhysRevB.99.094405}, or in-plane four-fold crystal anisotropy \cite{10.1063/1.4936994}. In the absence of intrinsic anisotropy, a piezoelectric stressor \cite{ferromagnetism} or the lattice mismatch between the magnetic layer and the nonmagnetic substrate \cite{PhysRevB.86.144420,Shibata2015} can induce anisotropy constants with a wide range of tunability. The application of an electric field produces the second term and provides a direct external parameter to tune skyrmion helicity \cite{Yao_2020}, while the last term corresponds to the application of an external magnetic field gradient or the presence of a DMI term. 

Depending on the parameter regime, two fundamental qubit designs are identified based on eigenstates of either $\hat{S}_z$, quantized magnetic excitations of the perpendicular to the 2D plane magnetization component, or $\hat{\varphi}_0$ states with a well-defined helicity. Notably, the former type resembles the superconducting charge qubit based on the number of Cooper pairs, while the latter the flux qubit based on quantized magnetic flux in a superconducting loop \cite{doi:10.1146/annurev-conmatphys-031119-050605}. The eigenstates of both qubit designs are mapped on a simple physical basis and are therefore useful for quantum annealing \cite{McGeoch2014}. Analogous to the anharmonicity caused by the nonlinear inductance of superconducting Josephson junctions, the electric field produces the necessary nonlinearity, causing non-equidistant level spacing. 

Indeed, the operation of the skyrmion qubit relies heavily on the anharmonicity of the well potential, in order to individually address transitions between quantized levels with a distinct frequency. By tuning the external electric and magnetic fields, a relatively large anharmonicity can be achieved $\vert \omega_{ex}-\omega_q \vert >20\% \omega_q$, where $\omega_q$ is the qubit frequency and $\omega_{ex}$ the frequency of higher level transitions \cite{PhysRevLett.127.067201}. This implies, that although the qubit frequencies between neighboring qubits depend on the fabrication process precision, the system will not suffer from the frequency collisions expected in weakly anharmonic multi-qubit systems \cite{8614500}.

The lowest two states $\vert 0 \rangle$ and $\vert 1 \rangle$ of the skyrmion Hamiltonian $H$ span the computational space and naturally form a tight two-level system with an approximate Hamiltonian
\begin{align}
H_q = \frac{H_0}{2} \hat{\sigma}_z - \frac{X_c}{2}\hat{\sigma}_x \,,
\label{eq:qubitHam}
\end{align}
where the form of $H_0$ is given in terms of the original parameters and depends on the qubit design, and $X_c$ is the control parameter that corresponds to either the electric field or the magnetic field gradient. In Eq.~\eqref{eq:qubitHam}, $\hat{\sigma}_{z,x}$ refer to the Pauli spin operators. The qubit frequency is $\omega_q=\sqrt{H_0^2+X_c^2}$. Helicity qubits work close to the degeneracy point $h_1 S \mathcal{M} =-1$, where the helicity degree of freedom becomes dominant and can be considered as the discrete variable. In the limit of large potential strengths $V_i \gg 1$, the lowest two states are well described by symmetric and antisymmetric combinations of the two wave functions localized in each well (see Fig.~\ref{Fig:skyrmion_qubit} for a representation). The qubit can be in a superposition of these two macroscopic states with distinct helicities, a manifestation of the quantum mechanical behavior of a macroscopic system. Coherent tunneling between these degenerate states through the potential barrier lifts the degeneracy and results in an energy-splitting of the order of MHz, well decoupled from the spectrum of higher levels in the GHz regime. 

\subsection{Noise and Decoherence}

Skyrmion qubits are macroscopic in size and are expected to be sensitive to environmental noise. The interaction with uncontrolled degrees of freedom, both extrinsic and intrinsic, inevitably results in quantum decoherence, a key challenge to address in the practical development of qubit devices. It is a delicate matter to isolate the qubit from a perturbing environment, and desirable operation and unwanted perturbation (noise) go hand in hand. Important timescales for the qubit decoherence are the energy relaxation time $T_1$ and dephasing time $T_2$. The former quantifies the decay of the first excited state $\vert 1 \rangle$ to the ground state $\vert 0 \rangle$, while the latter corresponds to the time over which the phase difference between two eigenstates in a quantum superposition state becomes randomized. Both $T_1$ and $T_2$ are important to estimate the expected accuracy of quantum operations and can be evaluated by a weak coupling of the skyrmion qubit to the quantum environmental noise. 

\begin{figure}[t]
	\centering
	\includegraphics[width=1\linewidth]{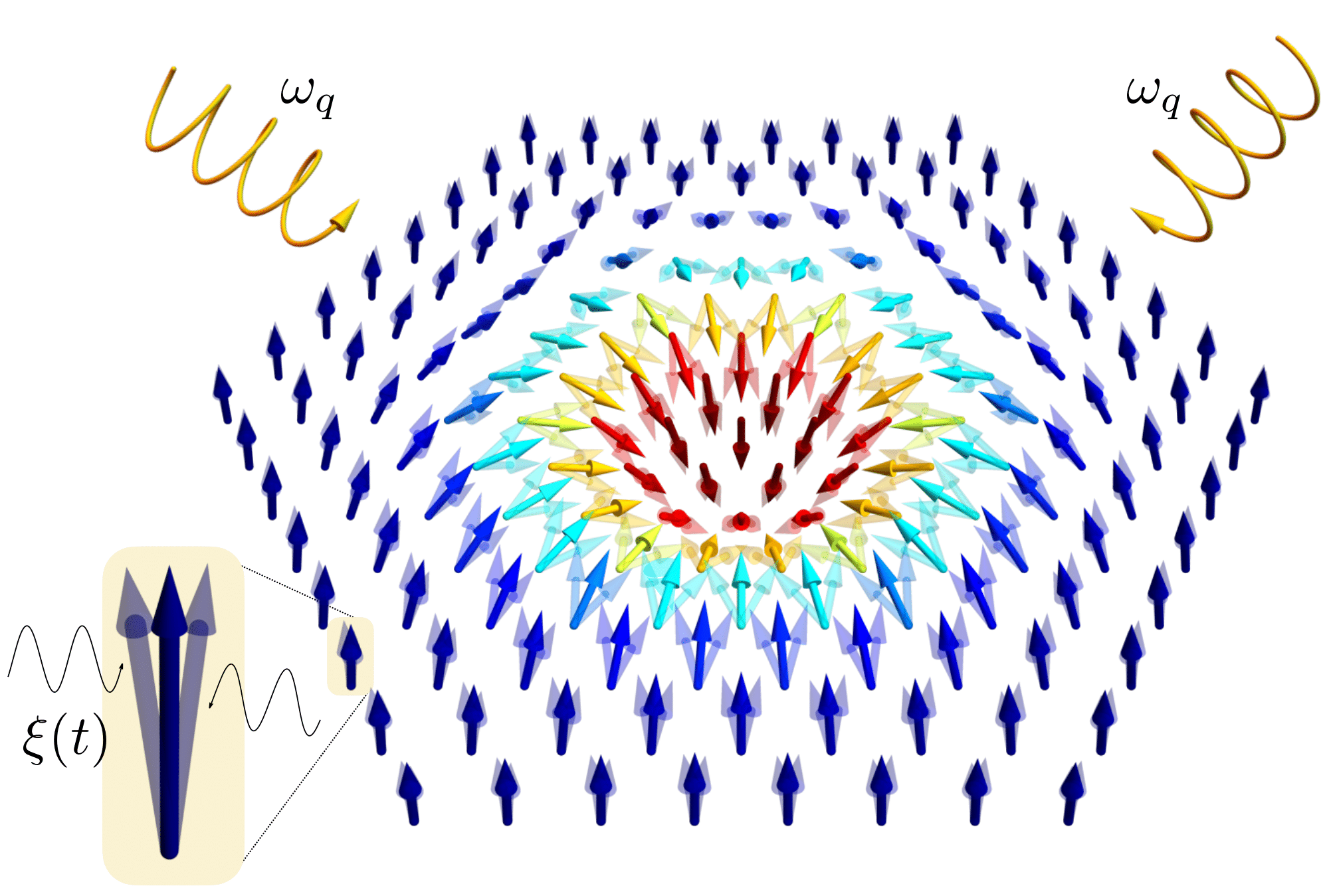}
	\caption{A magnetic skyrmion in the presence of noise gives rise to random fluctuating forces $\xi(t)$ (black arrows) acting on the skyrmion qubit, leading to decoherence. Driving the system with microwave magnetic fields (yellow arrows) at the qubit transition frequency $\omega_q$ can be used for single-qubit operations.}
	\label{Fig:skyrmion_driven}
\end{figure}

Decoherence times are estimated starting from the magnetization dynamics encoded in the Landau-Lifshitz-Gilbert equation (LLG) \cite{landau1980course,1353448}. Possible damping mechanisms for the collective coordinate of helicity are parameterized by a phenomenological velocity-dependent Ohmic term $\alpha_{\varphi_0} \dot{\varphi}_0$ induced by the coupling of the skyrmion to other unspecified degrees of freedom \cite{PhysRevLett.127.067201}. Here $\alpha_{\varphi_0}$ is a constant proportional to the Gilbert damping $\alpha$. Dissipation terms are accompanied by random fluctuating forces $\xi_i$, entering the quantum Hamiltonian as $H_q=-\frac{1}{2}\omega_q \sigma_z + \xi_{\perp} \sigma_{\perp} + \xi_z \sigma_{z}$, where $\sigma_{\perp}$ denotes the transverse spin components $\sigma_{x,y}$ (see Fig.~\ref{Fig:skyrmion_driven} for a visualization). Noise sources are conveniently described by their quantum spectral density $\langle \xi_i(t) \xi_j(t') \rangle = \delta_{ij} S_{ij}(t-t')$, linked to the dissipative terms via the fluctuation-dissipation theorem. Using the standard Bloch-Redfield \cite{PhysRev.91.1085,PhysRev.105.1206} picture of two-level system dynamics, both $T_1$ and $T_2$ are estimated to be in the microsecond regime, thus comparable to early measurements of the flux superconducting qubit and two orders of magnitude larger than the Cooper pair box \cite{Kjaergaard2020}. Here we assume an ultra-low Gilbert damping $\alpha=10^{-5}$ and low operational temperature $T=100$ mK. 

Although the total noise experienced by the skyrmion has multiple origins, the phenomenological LLG equation leaves the microscopic details of the skyrmion-environment coupling unspecified. Decoherence rates can be estimated more precisely by deriving a microscopic description of the properties of skyrmions coupled to quasiparticles naturally excited due to thermodynamics, i.e. itinerant electrons, phonons, and magnons. In magnetic insulators and at low enough temperatures, the spin degrees of freedom are the prominent source of noise. A microscopic treatment of the dynamics of the skyrmion center-of-mass coupled with a bath of magnon-like quantum excitations predicts super-Ohmic damping terms at temperatures below the magnon gap, which remain finite down to zero temperature due to the quantum nature of the bath \cite{PhysRevX.7.041045}. Damping terms are accompanied by random stochastic fields with a colored auto-correlation function \cite{PhysRevB.100.134404} and have a less detrimental effect on the quantum behavior of mesoscopic systems when compared with Ohmic-like noises with short correlation times \cite{CALDEIRA1983374,PhysRevB.46.5392}. 

In metallic ferromagnets, conduction electrons have strong effects on the skyrmion motion, as they generate (non)-adiabatic spin-transfer torques and damping terms \cite{TATARA2019208}. Within a field-theoretical treatment, the non-adiabatic contribution of conduction electrons gives rise to quantum inertia terms calculated microscopically \cite{PhysRevB.92.184410,PhysRevB.101.054407,2302.04220}. Finally, an inertia mass has been computed exactly as the result of the skyrmion-phonon interaction within a toy model of the magnetoelastic coupling\cite{PhysRevB.102.060404}. Generalization of the above theoretical approaches to include the coupling of the skyrmion helicity $\varphi_0$ to quasiparticle noise sources is straightforward and will result in a more accurate estimation of the decoherence times. 

Other types of noise include local fluctuating electric and magnetic fields, nuclear spins and dipolar interactions, structural defects in the sample and at interfaces, contributions from the leads connected to the solid-state devices, as well as systematic noise associated with the microwave pulse resonator or readout circuits. The degree to which a qubit is influenced by these noise sources is related to the qubit’s susceptibility, which in turn is determined by fabrication processes, cryogenic engineering, and electronics design. Longer coherence times can be achieved by developing a design strategy aiming to operate at the skyrmion qubit's optimal regime.

\subsection{Universal Quantum Computing}

Quantum algorithms are implemented by a small set of single-qubit and two-qubit unitary operations, the basis of the future skyrmion quantum processors. Single-qubit gates $U_{i}(\theta)$ rotate an arbitrary Bloch vector at a certain angle $\theta$ about a particular axis $i=\{x,y,z\}$, while two-qubit entangling operations are generally conditional gates involving two qubits. A complete single-qubit gate set supplemented with an entangling two-qubit operation suffices for universality, in the sense that all unitary operations on arbitrarily many bits can be expressed as compositions of these gates \cite{PhysRevA.52.3457}. A common universal quantum gate set is $\mc{G}_0=\{U_x(\theta),U_y(\theta),U_z(\theta), U_{\mbox{\scriptsize ph}} (\theta), U_{\mbox{\scriptsize CNOT}} \}$, where $U_{\mbox{\scriptsize ph}} (\theta) = e^{i \theta} \mathbb{1}$ applies a phase $\theta$ and $U_{\mbox{\scriptsize CNOT}}$ flips the state of the second qubit conditioned on the first qubit being in state $\vert 1 \rangle$. Another universal gate set is $\mc{G}_1=\{ U_H, U_S, U_T, U_{\mbox{\scriptsize CNOT}} \}$, where the $U_H$ Hadamard gate performs a $\pi$ rotation about the $(x+z)/2$ axis, and the $U_S$ ($U_T$) rotates the qubit state by $\pi/2$ ($\pi/4$) around the $z$ axis. Any other single-qubit gate can be approximated using only single-qubit gates from $\mc{G}_1$ \cite{DawsonNielsen}. 

\subsubsection{Single-Qubit Gates}

Microwave magnetic field gradients with frequencies at the qubit transition $\omega_q$ can be used to drive single-qubit gates \cite{PhysRevLett.127.067201}. In particular, one can generate rotations around the $x$ and $y$ axis by controlling the coupling between the qubit states $\vert 0 \rangle$ and $\vert 1 \rangle$ using microwave pulses. The driven qubit Hamiltonian in the frame rotating with the qubit frequency reads $H_q=\Delta \omega_q/2 \sigma_z + \Omega f(t)/2 [\cos \phi \sigma_x + \sin \phi \sigma_y]$, where $\Delta \omega = \omega_q - \omega$ is the detuning frequency with $\omega$ the frequency of the microwave source, $\Omega$ depends on the external field amplitude, $f(t)$ is a dimensionless pulse-envelope function and $\phi$ the phase of the external drive. Rotation around the $x$ axis by an angle $\vartheta(t)=-\Omega \int_0^t f(t') dt'$, i.e. $U_{x}(\vartheta(t)) = e^{-i \vartheta(t) \sigma_{x}}$, is performed using resonant driving $\omega=\omega_q$ and in-phase pulses $\phi=0$. Analogously, $U_{y}(\vartheta(t))$ is achieved with out-of-phase pulses $\phi=\pi/2$. 

Rotations around the remaining $z$ axis are generated by adjusting the phase $\phi$ of the drive. These are known as virtual zero-duration $U_z$ gates and correspond to adding a phase offset to the drive field for subsequent $U_x$ and $U_y$ gates \cite{PhysRevA.96.022330}. In addition to microwave gates, a time-dependent electric field protocol generates the $U_z(\theta)$ gate, while a time-dependent spin current performs rotations of the Bloch vector around the $x$ axis and generates the $U_x(\theta)$ gate \cite{PhysRevLett.130.106701}. It becomes apparent that the gates $U_{x,y,z}$ are simple to implement and are natively available in a skyrmion quantum processor. Any SU(2) gate is constructed by combining $U_z$ and $U_x$ gates \cite{PhysRevA.96.022330}. As an example, the Hadamard gate is expressed as $U_H=U_z(\pi/2)\cdot U_x(\pi/2)\cdot U_z (\pi/2)$. 

\begin{figure}[b]
	\centering
	\includegraphics[scale=0.35]{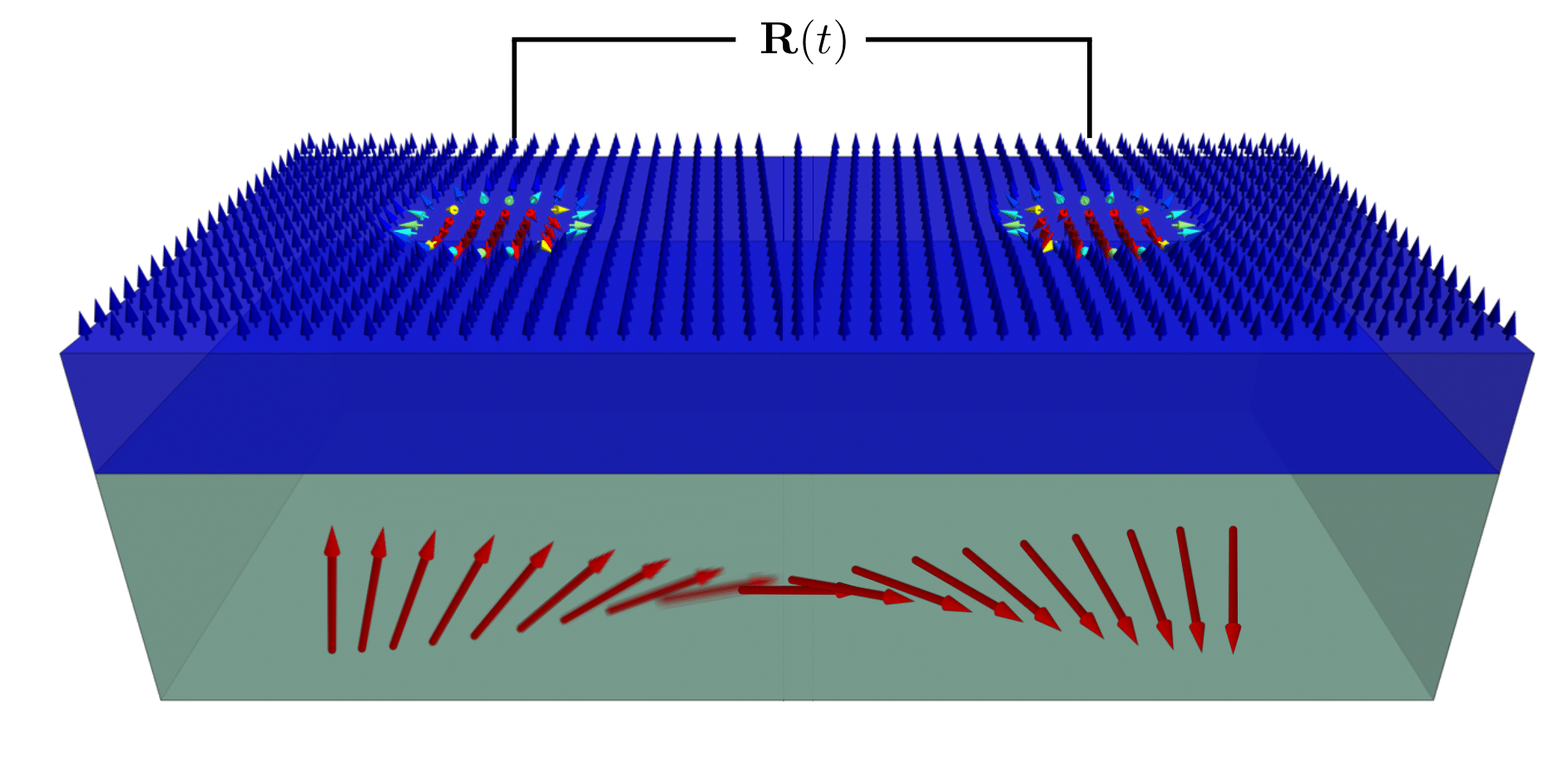}
	\caption{Illustration of a two-skyrmion coupling scheme. Two magnetic skyrmions corresponding to two distinct qubits are placed in a nanowire (blue), separated by a distance $\mb{R}(t)$. Two-qubit gates are implemented by a time-dependent external field protocol controlling $\mb{R}(t)$. Alternatively, long-distance two-qubit entanglement is possible via magnon modes (red arrows) excited in low-damping magnetic materials (green).}
	\label{Fig:two_skyrmion}
\end{figure}

\subsubsection{Two-Qubit Gates} 

Alongside arbitrary single-qubit gates, the universality of particular quantum hardware is demonstrated by outlining methods for generating two-qubit gates. Among them, the controlled-NOT (CNOT) gate and the controlled phase (CZ) are commonly used points of reference in the creation of quantum circuits. CNOT flips the state of the target qubit and CZ applies a Z gate to the target qubit, conditioned on the control qubit being in state $\vert 1 \rangle$. 

In magnetic materials, the interlayer exchange interaction provides a straightforward two-skyrmion coupling scheme \cite{PhysRevLett.127.067201}. Bilayers inherently give rise to interacting skyrmions across various interfaces with distinct helicity-dependent dynamical phenomena \cite{Koshibae2017,Diaz2020}. The ferromagnetic interaction term $-J_{\mbox{\scriptsize int}} \int_{\mb{r}} \mb{m}_1(\mb{r}) \cdot \mb{m}_2(\mb{r})$, with $i=\{1,2\}$ the layer index, translates into interaction between the two skyrmion helicities $H_{\mbox{\scriptsize int}}= - J'_{\mbox{\scriptsize int}} \cos (\varphi_1-\varphi_2)$ and a two-qubit Hamiltonian with both transverse and longitudinal couplings, $H_{\mbox{\scriptsize int}}= - \mc{J}^x_{\mbox{\scriptsize int}} \sigma_\perp^1 \sigma_\perp^2 - \mc{J}^z_{\mbox{\scriptsize int}} \sigma_1^z \sigma_2^z$. The interlayer interaction $J_{\mbox{\scriptsize int}}$ is controlled by either a nonmagnetic insulating spacer, such that $J_{\mbox{\scriptsize int}}$ decays exponentially with the spacer thickness, or a metallic spacer, such that the exchange interaction oscillates and can change sign \cite{Koshibae2017}. Alternatively, $J_{\mbox{\scriptsize int}}$ is tuned by varying the distance between the layers. 

The two-qubit Ising interaction $\mc{J}^z_{\mbox{\scriptsize int}} \sigma_1^z \sigma_2^z$ yields the CNOT and CZ gates directly using one two-bit operation \cite{PhysRevA.67.032301} under a time-dependent skyrmion bilayer interaction protocol \cite{PhysRevLett.130.106701}. The two skyrmion-qubit Heisenberg interaction $\mc{J}_{\mbox{\scriptsize int}} (\sigma_1^x \sigma_2^x+\sigma_1^y \sigma_2^y+\sigma_1^z \sigma_2^z)$ corresponds to the SWAP operation that exchanges the states of two qubits \cite{PhysRevA.67.032301} and can generate the CNOT gate by two two-bit operations \cite{PhysRevA.67.032301}. Two-qubit gates in magnetic domain wall qubits are implemented using the coupled dynamics of the DW position and chirality \cite{2212.12019}, similar to the coupling of the skyrmion helicity to the translational motion \cite{Zhang2017}. 

In addition to the direct bilayer magnetic interaction, spatially separated skyrmion qubits can be entangled using their interaction to delocalized degrees of freedom. Quasiparticle-mediated entanglement proposals for hybrid quantum systems provide a high degree of control over interactions between solid-state magnetic qubits. Long-distance spin-qubit coupling via magnon modes in ferromagnets can be achieved under realistic experimental conditions \cite{PRXQuantum.2.040314} (see Fig.~\ref{Fig:two_skyrmion} for a visualization of the entanglement platform). Photon-mediated magnon-magnon coupling between magnetoelectrically active skyrmion excitations in a cavity can be used to generate entanglement between magnon qubits \cite{PRXQuantum.3.040321}. These schemes are directly extended to the skyrmion qubit platform and allow the design of optimal protocols for gate synthesis \cite{PhysRevLett.88.237902}. 

\subsection{Qubit Readout}

Quantum computing requires coupling quantum systems to external instruments for control and readout. We need fast and accurate qubit readout and hardware that facilitates high qubit connectivity. The challenge, however, is to control and measure qubits while prohibiting unwanted interactions with their environment and scaling them up to larger systems. One architecture for realizing quantum processors can be skyrmion qubits coupled to coplanar waveguide resonators for readout and control of one another. The resonator frequency is qubit-state dependent, and thus, through probing this resonator, it becomes possible to discern the qubit state. Applying a readout tone at the qubit frequency, we can measure the state-dependent phase shift of the reflected readout tone. Detectors with a few electrodes, high readout fidelity, and speed are required to discern the skyrmion qubit state in a measurement time of under a few microseconds. In other words, a timescale shorter than the decoherence time, $T_{1,2}$, enables the implementation of error correction.

Before devices are developed, however, it is important to demonstrate experimentally the quantization of helicity. Helicity-qubit states represent two distinct skyrmion configurations with helicity values located at the two minima of the double-well potential \cite{PhysRevLett.127.067201}. Microwave techniques are especially suitable to characterize materials exhibiting non-collinear textures. Resonant modes detecting helicity shifts can be traced as a function of frequency, magnetic, and electric fields. Resonance probes can detect eigenstate transitions in individual skyrmions, either by magnetic resonance force microscopy (MRFM) \cite{Midzor2000,Arima_2015}, or by using a magnetic force microscopy (MFM) tip as a local microwave probe \cite{SAPOZHNIKOV2022169043,Hug2021,Midzor2000,Arima_2015,PhysRevLett.123.167201,Satywali2021}. 

Ferromagnetic Resonance (FMR) is a useful method to distinguish helicities, and detect gyration and breathing modes of skyrmions \cite{PhysRevLett.123.167201,Satywali2021}, with a quantized level shift manifesting as a discontinuity in the resonance frequency in the millikelvin regime. For a skyrmion radius of only a few nanometers, it is challenging to probe single skyrmions, and therefore collective excitations need to be analyzed in a magnetically homogeneous material. A complementary approach is circular X-ray polarization combined with a microwave drive to track Rabi oscillations. Furthermore, single nitrogen-vacancy in diamond tips has been shown to detect skyrmion helicities \cite{Dovzhenko2018,PhysRevB.105.144430}, although a spatial resolution of less than 10 nm remains a challenge \cite{Marchiori2021,doi:10.1021/nl0719271,Pezzagna_2010,cryst7050124}. At a practical level, spin helicity in itinerant non-collinear magnets can be controlled with the use of magnetic fields and electric currents.  Hence, a convenient read out is possible by second-harmonic resistivity, and via tunneling magnetoresistance at a segment of the skyrmion,
potentially enabled by advances in nanofabrication capabilities \cite{Jiang2020,LimaFernandes2022}.

\section{Future Perspectives}

Finding and stabilizing new spin configurations in materials is one of the fundamental goals of condensed matter physics. A plethora of compounds exhibit a remarkable range of magnetic phases including nano-skyrmions that are of direct relevance to quantum technology. While the traditional approach has been to search for such phases within naturally occurring compounds, advances in the angstrom-scale layer-by-layer synthesis of multi-element compounds for materials by design have taken the approach to a new level of power and sophistication. This grants access to a controlled terrain of materials engineering and electronic operations at the atomic scale, a particularly appealing hunting ground for new physics and targeted applications. Already, the potential of certain nano-skyrmions for quantum hardware has triggered proposals for device and circuit configurations. Combined with the availability of materials hosting nano-skyrmions with distinct helicity, this engenders fundamentally new helicity-based operations in a convenient parameter space, with much promise for a skyrmion-based qubit. 

\subsection{Candidate Materials}

Candidate materials for a skyrmion qubit architecture need to satisfy certain criteria. A variable-helicity skyrmion qubit may be materialized in frustrated magnets with a suitable crystal structure and ultra-low Gilbert damping of 10$^{-5}$ to 10$^{-4}$. Insulating materials are preferable for placing electronic gates in direct contact with the magnetic film,  but so far are rare hosts of sub-10 nm skyrmions at low temperatures. On the other hand, there is already encouraging evidence in metallic frustrated magnets with low and tunable charge carrier density. Here, an insulating layer placed between the gates and the magnet would allow the desired nonvolatile control. 

Promising candidates include Gd$_{3}$Ru$_{4}$Al$_{2}$ and Gd$_{2}$PdSi$_{3}$ \cite{doi:10.1126/science.aau0968,Chandragiri_2016, Hirschberger2019}. Gd$_{2}$PdSi$_{3}$ features frustrated triangular lattice planes of magnetic Gd ions interleaved with a non-magnetic PdSi$_{3}$ honeycomb lattice. Other rare earth intermetallic magnets share the same space group, including RGa$_{2}$ (where R is a rare earth element) and ErSi$_{2}$ \cite{BORAGNO1994515,PhysRevB.56.15171}. The weak ferromagnet TbGa$_{2}$ \cite{AUNEAU1995351} shows signatures indicative of a non-collinear magnetic ground state \cite{doi:10.1126/science.aau0968,PhysRevLett.108.017206,Hirschberger2019}. NiGa$_{2}$S$_{4}$ is a Mott insulator built on a triangular lattice \cite{PhysRevB.78.180404,doi:10.1126/science.1114727,Nakatsuji_2007}, with a non-collinear antiferromagnetic ground state \cite{doi:10.1126/science.1114727,PhysRevB.77.054429}.

$\alpha$-NaFeO${_2}$ is also potentially promising although it is not yet clear whether its short-range spin correlations support a skyrmion lattice \cite{PhysRevLett.108.017206}. Bloch-type skyrmionic bubbles with degenerate $\pi$/2 and -$\pi$/2 helicities have been reported to stabilize in Fe$_{3}$Sn$_{2}$  \cite{Hou2019,Hou2020}. Here, both macroscopic states can be accessed via a reversible process that overcomes the energy barrier by applying an electric or magnetic field. CaBe$_{2}$Ge$_{2}$ is another interesting example. Although it is centrosymmetric in character with a layered structure, inversion symmetry is broken locally in the middle of the two layers \cite{Lin2021}, inducing local DMI and leading to a stable skyrmionic crystal lattice \cite{Lin2021}. MnInP$_{2}$Te$_{6}$ is predicted to stabilize a Bloch-type skyrmion lattice according to first-principles calculations and Monte-Carlo simulations \cite{Du2023}. Whereas SrFeO$_{3}$ has been shown to host skyrmions with a triple-q helical spin modulation \cite{PhysRevB.101.134406}. 

These are encouraging examples towards the identification of suitable materials for hosting skyrmions for quantum operations. We expect further work in materials engineering assisted by machine learning to unlock more compounds \cite{PMID:36262309,PhysRevB.100.174408,PhysRevB.98.174411}. A major challenge, however, is developing devices on high-quality epitaxial thin films, which may take time to reach optimal conditions for applications on an industrial scale. Meanwhile, thin lamellae can be cut from single crystals using a focused ion beam, following a similar routine to the sample preparation employed in transmission electron microscopy. Metallic top-gates and/or coplanar waveguides can be lithographically defined on hexagonal boron nitride (hBN) for electric field-tuned measurements.

\subsection{Noise Mitigation}

The interaction of the skyrmion qubit with the environmental degrees of freedom, however, leads to the collapse of the qubit’s superposition into one definite state \cite{PhysRevB.97.064401}. In practice, a skyrmion qubit is disrupted not only by interactions between its spin-whirls and the underlying magnet from which they emerge but also by the fluctuation of electromagnetic fields near the skyrmions. These forms of noise need to be combated with improved sample crystallinity, reduced damping, and dynamical decoupling from the environment by a set of pulses \cite{PhysRevLett.83.4888} in the multi-qubit gate \cite{vanderSar2012}.

Although magnetic thin films already exhibit the desired ultra-low Gilbert damping at room temperatures \cite{Soumah2018,Hauser2016,Qin2017,Chang2014}, the dependence of the damping coefficient on film thickness and temperature, especially at the sub-Kelvin qubit operational regime, is still not fully understood \cite{Guo2022,Haspot2022,Jin2019}. Furthermore, for better coherence times, it is imperative to focus on the development of cleaner magnetic samples and interfaces, without trading off qubit anharmonicity and scalability. High-quality layer-by-layer growth of heterostructures of skyrmion hosting perovskites such as SrFeO$_3$ is already possible by molecular beam epitaxy \cite{101063} and pulsed laser deposition \cite{CHANG2010621}. Similar progress is expected to occur in other skyrmion hosts. Indeed, controlled layer-by-layer deposition methods are more promising than the sputtering techniques commonly employed in the field of skyrmionics, as they offer the potential for realizing low-defect heterostructures. 

However, not just the magnetic thin film's quality is of critical importance. In the simplest skyrmion qubit platform, interlayer exchange interaction couples qubits via a nonmagnetic dielectric spacer, and logical states are adjusted by electric fields. Dielectrics with low loss at microwave frequencies are therefore crucial for high-coherence solid-state quantum computing platforms. Dielectric spacers between skyrmions are typically amorphous oxides with structural defects and their microscopic nature remains to be fully understood. However, hexagonal boron nitride (hBN) thin films offer a complementary platform to be suitably integrated into skyrmion qubit architectures \cite{Wang2022}. Here, the microwave loss tangent of ultra-thin films of hBN is at most in the mid 10$^{-6}$ range in the low-temperature regime. This is promising for building high-coherence quantum circuits with a substantial reduction of unwanted qubit cross-talk.

Besides strategies to mitigate quantum decoherence based on material engineering, alternative approaches built upon pulsed driving fields applied to the qubit might prove crucial. Dynamical decoupling \cite{PhysRevA.58.2733} relies on a sequence of pulses inducing fast qubit flips that average out environmental fluctuations at a specific frequency. It emerges as a particularly encouraging strategy for addressing the challenges of decoherence, as it has been shown to improve coherence times \cite{Bylander2011,101126} and can be integrated with quantum gates for a standard hybrid system \cite{vanderSar2012}. Alternatively, for magnetic systems, monochromatic \cite{RevModPhys.92.015004} and polychromatic \cite{Joos2022} driving of the spin bath enhances qubit coherence times. Thus, for skyrmions in magnetic insulators where decoherence is dominated by magnetic noise,  resonance frequency quantum control techniques can be employed on single qubits as well as on qubits linked up to form logic gates, to achieve optimal dynamical decoupling and improve coherence times.

\subsection{Device Architecture}

The integration of magnetic skyrmions into scalable quantum computing solid-state devices remains a challenge yet to be addressed. Utilizing a fundamental bottom-up approach in the design and engineering of quantum hardware involves tuning the skyrmion structure to produce the desired physical behavior. Positional control can be achieved by placing magnetic skyrmions in confined geometries in nanostructures \cite{Boulle2016,PhysRevApplied.11.024064,doi:10.1073/pnas.1600197113}. Alternatively, defect engineering can offer precise energy landscapes to control the skyrmion trajectory and position \cite{LimaFernandes2018,Arjana2020}. Depending on the quantum architecture dimensionality, skyrmions are stabilized in 1D and 2D arrays of nanodomes with useful properties arising from the curved geometry \cite{Tejo2020}. Eventually, skyrmion qubits can be manufactured in large arrays using the lithographic technology employed in microelectronics. For true quantum computing, however, skyrmion qubits will have to link up in large arrays of memory units and logical gates, leading to scalability concerns. 

The basic requirement for hardware here is the coupling of multiple skyrmions stabilized at ambient conditions and separated at the nanoscale, a feat for which encouraging results have already been reported using semiconductor-based capabilities \cite{PhysRevApplied.11.024064}. Skyrmion-skyrmion interaction in 2D magnetic films \cite{Zhang2015,Capic_2020} or bilayers \cite{Koshibae2017,Zhang2016} is sufficiently understood. Electric fields emerge as a new, powerful tool for a current-free low-power control of skyrmion dynamics. A reversible and reproducible skyrmion nucleation and annihilation mechanism has been discovered in a Pt/Co/oxide trilayer system \cite{Schott2017,Hsu2017} under electric field gating. The dynamic control of skyrmion helicity has been reported using voltage gating \cite{doi:10.1021/acs.nanolett.8b01502}. Finally, chiral skyrmions in multilayers X/CoFeB/MgO are manipulated under the influence of a magnetic field gradient using scanning tunneling microscopy (STM) tip \cite{Casiraghi2019}. Thus, the components of the suggested skyrmion qubit platform \cite{PhysRevLett.127.067201} are experimentally feasible with current technology.

\subsection{New Qubit Designs and Outlook}

The goal of this Perspective is to illustrate the potential magnetic skyrmions hold to  
exhibit controllable quantum behavior at the macroscale, allowing the implementation of quantum devices. We focused primarily on the conceptual framework and basic elements for the description of phenomena based on the quantum degree of skyrmion helicity.  Domain walls with opposite chirality can also serve as the fundamental building blocks of a quantum computer \cite{PhysRevB.97.064401,2212.12019} with similar decoherence time in the microsecond regime and operational frequency of a few GHz.  Quantum phenomena within magnetic textures are crucial yet largely unexplored, presenting a promising avenue for further investigation and discovery. However, the relevance of topological magnetic textures to quantum technologies lies beyond gate-based quantum computing. Foremost, although magnetic skyrmions can execute specific tasks within a quantum processor, devices with multitasking capabilities rely on the coherent interaction of skyrmions with other degrees of freedom with complementary functionalities. In particular, hybrid quantum systems based on skyrmions coupled to microwave and optical photons, atoms, and individual electron and nuclear spins \cite{Clerk2020}. 

Moreover, magnetic skyrmions might prove applicable to unconventional quantum information processing, including quantum reservoir computing based on quantum noise \cite{PhysRevApplied.8.024030,PhysRevApplied.14.024065}, adiabatic quantum computation \cite{quant-ph/0001106,doi:10.1126/science.1057726}, or methods that combine the advantages of different computational approaches \cite{Barends2016}. Skyrmions have been proposed for topological quantum computation, as they induce topological superconductivity when proximized with conventional superconductors \cite{PhysRevB.93.224505,Garnier2019}, forming skyrmion-vortex pairs hosting Majorana bound states (MBS). The experimental coupling of chiral magnetism and superconductivity \cite{PhysRevLett.126.117205} as well as the numerical demonstration of the non-Abelian statistics of MBS in this system \cite{PhysRevB.105.224509} represent encouraging results in the pursuit of a scalable topological quantum computer.

In practice, the underexplored quantum platform involving topological magnetic textures reveals characteristics with the potential to significantly enhance the progress of quantum technologies. Skyrmion qubits depict the nascent connection between quantum applications and spin topology, offering exciting prospects for generating and preserving quantum information through magnetic quantum states. 

\section{ACKNOWLEDGMENTS}
Christina Psaroudaki is an \'{E}cole Normale Sup\'{e}rieure (ENS) -Mitsubishi Heavy Industries (MHI) Chair of Quantum Information supported by MHI. The work in Singapore was supported by the National Research Foundation (NRF) Singapore Competitive Research Programme NRF-CRP21-2018-0001 and the Singapore Ministry of Education (MOE) Academic Research Fund Tier 3 Grant MOE2018-T3-1-002.

\nocite{*}
\bibliography{APL_SkyrmionQubit}
\end{document}